\documentclass[fleqn,twoside]{article}
\usepackage[headings]{espcrc2}

\readRCS
$Id: espcrc2.tex,v 1.2 2004/02/24 11:22:11 spepping Exp $
\usepackage{graphicx}
\usepackage[figuresright]{rotating}

\newcommand{\AmS}{{\protect\the\textfont2
  A\kern-.1667em\lower.5ex\hbox{M}\kern-.125emS}}

\title{Superconductivity Near a Quantum Critical Point in Ba(Fe$_{1-x}$Co$_{x}$)$_{2}$As$_{2}$}

\author{K. Ahilan\address[MCSD]{Department of Physics and Astronomy, McMaster University, Hamilton, Ontario L8S4M1, Canada},
        F. L. Ning\addressmark,
         T. Imai\addressmark\address{Canadian Institute for Advanced Research, Ontario M5G1Z8, Canada}\thanks{Research at McMaster was supported by NSERC, CFI, and CIFAR},
        A.S. Sefat\address[MCSD]{Materials Science and Technology Division, Oak Ridge National Laboratory, TN 37831, USA},
        M. A. McGuire\addressmark,
        B. C. Sales\addressmark,
        and
        D. Mandrus\addressmark[MCSD]\thanks{Research at ORNL was sponsored by Division of Materials Sciences and Engineering, Office of Basic Energy Sciences, U.S. Department of Energy.}
        P. Cheng\address[MCSD]{National Laboratory for Superconductivity, Institute of Physics and Beijing National Laboratory for Condensed Matter Physics, Chinese Academy of Sciences, Beijing 100190, China}, 
        B. Shen\addressmark,
        H.-H Wen\addressmark\thanks{Research at Beijing was supported by NSF, the Ministry of Science and Technology of China, and the Chinese Academy of Sciences.}, 
        }
       
\runtitle{Superconductivity Near a Quantum Critical Point in Ba(Fe$_{1-x}$Co$_{x}$)$_{2}$As$_{2}$}
\runauthor{K. Ahilan}

\begin{document}

\begin{abstract}
We will examine the possible link between spin fluctuations and the superconducting mechanism in the iron-based high temperature superconductor Ba(Fe$_{1-x}$Co$_{x}$)$_{2}$As$_{2}$ based on NMR and high pressure transport measurements.
\vspace{1pc}
\end{abstract}

\maketitle

\section{INTRODUCTION}

Over the last year, Co-doped BaFe$_{2}$As$_{2}$ \cite{Sefat} has emerged as an ideal platform for detailed investigation into the physical properties of iron-based high temperature superconductors.  The advantages of electron-doped Ba(Fe$_{1-x}$Co$_{x}$)$_{2}$As$_{2}$ are manyfold.   First and foremost, it is relatively straightforward to grow homogeneous single crystals \cite{Sefat}.  These single crystals allowed us to conduct a systematic NMR  \cite{Ning2,Ning1,Ning3,Ning4} and transport measurements \cite{Ning2,Ahilan2,Ahilan3} throughout a broad range of the phase diagram \cite{Ning2}.   Availability of high quality single crystals also led many other researchers to concentrate their efforts on investigating Ba(Fe$_{1-x}$Co$_{x}$)$_{2}$As$_{2}$.  Thanks to these concerted efforts, we can compare experimental results obtained by different techniques and build a comprehensive physical picture.  Furthermore,  the existence of an overdoped non-superconducting metallic regime \cite{Ni} allows us to investigate the fate of spin fluctuations when overdoping suppresses superconductivity \cite{Ning4}.  

In this invited paper, we will provide a perspective on the physical properties of Ba(Fe$_{1-x}$Co$_{x}$)$_{2}$As$_{2}$ focusing on two key issues.   First, we will extend our earlier transport measurements in ambient and applied pressure \cite{Ning2,Ahilan2,Ahilan3}, establish a new complete electronic phase diagram of Ba(Fe$_{1-x}$Co$_{x}$)$_{2}$As$_{2}$ under pressure of 2.4~GPa up to the overdoped regime, and discuss its implications.  Second, we will also deduce the temperature dependence of antiferromagnetic spin fluctuations (AFSF) in optimally doped Ba(Fe$_{0.92}$Co$_{0.08}$)$_{2}$As$_{2}$ based on a phenomenological two component analysis of our $^{75}$As NMR data \cite{Ning1}, and explain why AFSF may be the glue of superconducting Cooper pairs. 

\section{PHASE DIAGRAM}
In Fig.1, we reproduce the electronic phase diagram of Ba(Fe$_{1-x}$Co$_{x}$)$_{2}$As$_{2}$ in ambient pressure $P=0$ reported first by Ning et al. \cite{Ning2} with newer data points for the overdoped region $x \geq 0.1$ \cite{Ning4}.  A striking aspect of the phase diagram is that the superconducting dome is adjacent to an underdoped region with magnetically ordered ground states.  Analogous proximity between superconducting and magnetic phases has been encountered in many unconventional superconductors in the past, including the high $T_c$ cuprates. 

\begin{figure}[t]
\begin{center}
\includegraphics[width=7cm]{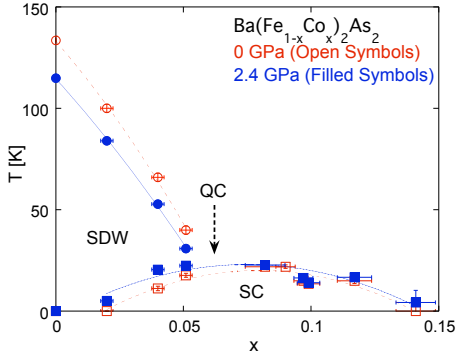}
\end{center}
\caption{The phase diagram of Ba(Fe$_{1-x}$Co$_{x}$)$_{2}$As$_{2}$ in ambient pressure (open symbols) and 2.4~GPa (filled symbols).  
}
\label{f1}
\end{figure}  

Also shown in Fig.1 are the magnetic phase transition temperature $T_{SDW}$ and the superconducting $T_c$ under 2.4~GPa of hydrostatic pressure, as determined by resistivity measurements.  We refer readers to Ref. \cite{Ahilan2,Ahilan3} for the details of experimental procedures.  We note that this is the first report on the effects of pressure on $T_c$ in the overdoped region of Ba(Fe$_{1-x}$Co$_{x}$)$_{2}$As$_{2}$.  Application of 2.4~GPa pushes the SDW phase boundary toward smaller $x$, and extends the superconducting dome.  One can certainly argue that this is evidence for competition between superconductivity and magnetism.  However, such a viewpoint may be too simplistic, because suppression of the SDW ordered regime is accompanied by the creation of a new paramagnetic regime with enhanced paramagnetic spin fluctuations; the latter may cause the expansion of the superconducting dome to smaller values of $x$.

In passing, it is worth noting that $T_c$ {\it increases} by $\Delta T_{c}\sim 2$~K under pressure for both $x=0.12$ ($T_{c}^{mid} = 15.0$~K in $P=0$) and $x=0.14$ ($T_{c}^{onset} = 6.0$~K in $P=0$).  Although the pressure coefficient is only modest in the overdoped regime (e.g. $dT_{c}/dP = +0.71$~K/GPa for $x=0.12$), $dT_{c}/dP$ is always {\it positive} in the entire phase diagram.  Our finding is in contrast with the case of hole-doped K$_{x}$Sr$_{1-x}$Fe$_{2}$As$_{2}$, where $dT_{c}/dP$ changes its sign from positive to negative in  the overdoped regime \cite{Chu}.  The results for K$_{x}$Sr$_{1-x}$Fe$_{2}$As$_{2}$ were interpreted in the context of transfer of holes from K$_{x}$Sr$_{1-x}$ charge reservoir layers to FeAs layers under pressure.  In the present case, we are doping electrons directly into FeAs layers by substituting Fe with Co.

\section{SPIN FLUCTUATIONS}
The exact nature of the magnetically ordered ground state under the presence of Co is still somewhat controversial, but there is a consensus that the magnetic phase transition is second order in Ba(Fe$_{1-x}$Co$_{x}$)$_{2}$As$_{2}$ for $x >0$.  In fact, we observe a divergent signature toward $T_{SDW}$ in the temperature dependence of $1/T_{1}T$, i.e. the NMR spin-lattice relaxation rate $1/T_{1}$ divided by $T$ \cite{Ning2,Ning3}.   $1/T_{1}T$ probes the weighted \textbf{q}-integral in the Brillouin zone (B.Z.) of low frequency spin fluctuations at the NMR frequency $\omega_{o}/2\pi \sim 50$~MHz,  

\begin{equation}
1/T_{1}T \propto \sum_{{\bf q}\in B.Z.} {|A({\bf q})|^{2} \frac {\chi " ({\bf q}, \omega_{o})}  {\omega_{o}} },
\end{equation}
where $A({\bf q})$ is the hyperfine form factor, and $\chi " ({\bf q}, \omega_{o})$ is the imaginary part of the dynamical electron spin susceptibility.  The divergent behavior of $1/T_{1}T$ towards a magnetic phase transition signals the critical slowing down of spin fluctuations expected for second order magnetic  phase transitions.  In other words, near the SDW phase boundary, low frequency spin fluctuations are highly enhanced.  

In view of the proximity between the SDW and superconducting phases in Fig.1, a natural question  is if spin fluctuations are enhanced even in the normal metallic state above $T_{c} = 22$~K of the optimally doped Ba(Fe$_{0.92}$Co$_{0.08}$)$_{2}$As$_{2}$.   Our earlier NMR measurements answered this question \cite{Ning2,Ning1,Ning4}.  Fig.2 summarizes the key physical properties of Ba(Fe$_{0.92}$Co$_{0.08}$)$_{2}$As$_{2}$ \cite{Ning1,Ahilan2}.  Our $^{75}$As NMR data in Fig.2c indeed captured a clear signature of enhancement of $1/T_{1}T$ from $\sim 100$~K to $T_c$.  The enhancement of $1/T_{1}T$ toward $T_c$ is stronger when we apply the external magnetic field $B_{ext}$ along the ab-plane rather than the c-axis.  This is because the ab-plane components of the hyperfine magnetic fields transferred from Fe layers accidentally cancel out at $^{75}$As sites for commensurate antiferromagnetic wave vectors.  $1/T_{1}T$ probes spin fluctuations orthogonal to the quantization axis of nuclear spins, and the latter is along the direction of $B_{ext}$.  Accordingly, $1/T_{1}T$ with $B_{ext}//c$ is less efficient in capturing AFSF.

Since the divergent behavior of  $1/T_{1}T$ at $T_{SDW}$ for $x \leq 0.06$ arises from slowing of AFSF for the wave vector modes ${\bf Q_{AF}} =$~(${\pi}/a$,0) and  (0,${\pi}/a$), we can infer that the same (or similar) modes of AFSF near ${\bf Q_{AF}}$  are enhanced in the optimal superconducting composition toward $T_c$.  

\begin{figure}
\begin{center}
\includegraphics[width=11cm]{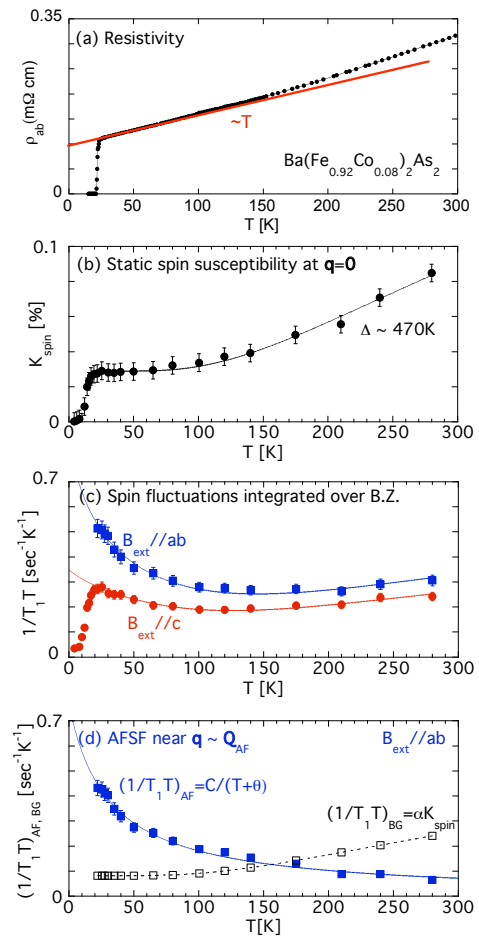}
\end{center}
\caption{Key physical properties of Ba(Fe$_{0.92}$Co$_{0.08}$)$_{2}$As$_{2}$ single crystal ($T_{c}=22$~K): (a) in-plane resistivity with a T-linear fit \cite{Ahilan2}. (b) Spin susceptibility deduced from the NMR Knight shift \cite{Ning1}. (c) Temperature dependence of low frequency spin fluctuations integrated over the entire B.Z., as measured by $^{75}$As NMR $1/T_{1}T$ \cite{Ning2}.  (d) Deconvolution of the results in (c) into the contribution from the antiferromagnetic ${\bf q} \sim {\bf Q_{AF}}$ mode of spin fluctuations $(1/T_{1}T)_{AF}$ and the background contribution $(1/T_{1}T)_{BG}$ (see main text).
}
\label{f2}
\end{figure}  

Another interesting point in Fig.2c is that the temperature dependence of $1/T_{1}T$ is not monotonic; the overall spin fluctuations integrated over the entire first B.Z.  {\it decrease} with temperature from 290~K down to about 100~K.  Furthermore, the temperature dependence of static uniform (${\bf q}={\bf 0}$) spin susceptibility deduced from the spin contribution to the NMR Knight shift $K_{spin}$ decreases with temperature, as shown in Fig.2b.  Comparison of Fig.2b and Fig.2c suggests that spin excitations are {\it suppressed} with decreasing temperature for a broad range of wave vector modes, including a region near the B.Z. center ${\bf q} = {\bf 0}$.  We recall that an analogous suppression of overall spin excitations was first observed in LaFeAsO$_{0.89}$F$_{0.11}$ ($T_{c}=28$~K) \cite{Ahilan1}, and more recently in stoichiometric FeSe ($T_{c}=9$~K) \cite{ImaiFeSe}, and hence this is probably a generic feature shared by iron-based superconductors.  Simultaneous suppression of both NMR Knight shift and $1/T_{1}T$ is generally called {\it spin gap} or {\it pseudo gap} behavior.  

Recently, inelastic neutron scattering measurements on Ba(Fe$_{0.925}$Co$_{0.075}$)$_{2}$As$_{2}$ also confirmed  the enhancement of AFSF toward $T_c$ \cite{Inosov}.  However, neutron data showed monotonous {\it increase} of AFSF all the way from 280~K down to $T_c$, without the initial decrease observed in Fig.2c from 290~K to 100~K.  The key to understanding the reason behind the apparent contradiction between NMR and neutron data above 100~K is that $1/T_{1}T$ reflects the integral of spin fluctuations over the entire B.Z., while the neutron data in Ref. \cite{Inosov} integrates only a peak located at ${\bf Q_{AF}}$.  We emphasize that NMR is very good at detecting small $\chi " ({\bf q}, \omega_{o})$ even if there are no pronounced peaks in the {\bf q} and/or $\omega$ space.  Combined with the neutron data, our results in Fig.2c suggest that  low energy spin excitations in a broad range of the wave vector space far from ${\bf Q_{AF}}$ are suppressed with decreasing temperature from 290~K down to 100~K. 

 To illustrate our point more clearly, we employ a phenomenological two component picture and assume that $1/T_{1}T = (1/T_{1}T)_{AF} + (1/T_{1}T)_{BG}$.  $(1/T_{1}T)_{AF}$ arises from AFSF with  ${\bf q}\sim {\bf Q_{AF}}$, and we further assume (for simplicity) that it obeys a Curie-Weiss law, $(1/T_{1}T)_{AF}= C/(T+\theta)$.   On the other hand, $(1/T_{1}T)_{BG}$ represents the contributions by the background of the dynamical susceptibility far from ${\bf Q_{AF}}$, which may have only mild ${\bf q}$ dependence and would be very difficult to observe using neutron scattering.  In view of the similarity of the temperature dependence between the overall $1/T_{1}T $ and $K_{spin}$ between 290~K and 100~K, it is reasonable to make a working ansatz, $(1/T_{1}T)_{BG} \propto K_{spin}$.  We showed earlier \cite{Ning1} that $K_{spin} = 0.027 + 0.29~exp(-\Delta/k_{B}T)$ with a pseudo gap $\Delta/k_{B} \sim 470$~K, as shown by a solid curve in Fig.2b.  The solid curves in Fig.2c represent the best fit of the data with the phenomenological two-component model, $1/T_{1}T = \alpha K_{spin} + C/(T+\theta)$, where $\alpha$, $C$, and $\theta$ are the free parameters.   We deduced the temperature dependence of $(1/T_{1}T)_{AF}$ as $1/T_{1}T - \alpha K_{spin}$ from the fit in Fig.2c,  and the results are presented in Fig.2d.  We found $\theta\sim31$~K for $B_{ext}//ab$.  Very small $\theta$ is consistent with a viewpoint that the optimally doped Ba(Fe$_{0.92}$Co$_{0.08}$)$_{2}$As$_{2}$ is in the vicinity of a quantum critical point \cite{Ning2}.  Also notice that our results of $\chi " ({\bf q}\sim{\bf Q_{AF}}, \omega_{o})$  deduced as $(1/T_{1}T)_{AF}$ in Fig.2d show almost identical behavior to the neutron scattering data  integrated near ${\bf q}\sim {\bf Q_{AF}}$ for energy transfer $\omega =3$~meV\cite{Inosov}. 

\section{CONCLUSIONS}

We have demonstrated that the optimal high $T_c$ superconducting phase Ba(Fe$_{0.92}$Co$_{0.08}$)$_{2}$As$_{2}$ exists in close proximity with magnetically ordered ground state, and that low frequency antiferromagnetic spin fluctuations are still enhanced near $T_{c}=22$~K.  Based on a phenomenological two-component model analysis, we also explained that the pseudogap like behavior above $\sim 100$~K arises from the suppression with temperature of a background spin susceptibility spread over a broad range of {\bf q} values away from  ${\bf Q_{AF}}$.  

The NMR results for the optimal superconducting phase alone do not necessarily prove that spin fluctuations are the {\it cause} of superconductivity.  One can, in principle, argue that  $T_c$ is as {\it low} as 22~K in Ba(Fe$_{0.92}$Co$_{0.08}$)$_{2}$As$_{2}$ because the residual antiferromagnetic spin fluctuations are disrupting the formation of Cooper pairs.   However, there are two pieces of strong evidence which support the idea that AFSF help the formation of Cooper pairs.  First, the suppression of superconductivity in the overdoped region above $x\sim 0.15$ is accompanied by that of AFSF \cite{Ning4}.  If AFSF tend to suppress $T_c$ for $x\sim 0.08$, the suppression of AFSF would have to enhance $T_c$ for $x>0.08$ instead.  Second, we also found in the related compound FeSe that the application of pressure enhances $T_c$ and AFSF simultaneously \cite{ImaiFeSe}.

\end{document}